\documentclass{iaus}
\usepackage{graphicx}

\def\apj{\textit{ApJ}}
\def\apjl{\textit{ApJL}}
\def\aap{\textit{A\&A}}
\def\pasp{\textit{PASP}}

\def\araa{\textit{ARAA}}
\def\pasa{\textit{PASA}}
\def\physrep{\textit{PhysRep}}
\def\aaps{\textit{A\&AS}}

\title[Observing compact disks inside pre-PNe with the VLTI] 
{Observing compact disks inside pre-PNe with the VLTI}

\author[S.N. Bright, O. De Marco, O. Chesneau, E. Lagadec, H. Van Winckel, B. J. Hrivnak]   
{S. N. Bright$^{1,2}$
O. De Marco$^{1,2}$
O. Chesneau$^3$
E. Lagadec$^4$
H. Van Winckel$^5$
B. Hrivnak$^6$
}

\affiliation{

$^1$Macquarie University Research Centre in Astronomy, Astrophysics \& Astrophotonics \\[\affilskip]

$^2$Macquarie University, NSW, Australia email: {\tt stacey.bright@mq.edu.au, orsola.demarco@mq.edu.au} \\[\affilskip]

$^3$Observatoire de la C\^{o}te d'Azur,  Nice, France email: {\tt Olivier.Chesneau@obs-azur.fr} \\[\affilskip]
$^4$European Southern Observatory, Garching, Germany email: {\tt elagadec@eso.org} \\[\affilskip]
$^5$Instituut voorSterrenkunde, Leuven, Belgium email: {\tt  Hans.VanWinckel@ster.kuleuven.be} \\[\affilskip]
$^6$Valparaiso University, Valparaiso, Indiana, USA email: {\tt Bruce.Hrivnak@valpo.edu}
}

\pubyear{2011}
\volume{283}  
\pagerange{??--??}
\jname{Planetary Nebulae: An Eye to the Future}
\editors{A.C. Editor, B.D. Editor \& C.E. Editor, eds.}
\begin{document}

\maketitle

\begin{abstract}
AGB stars appear to lose mass spherically, but many PNe resulting from the AGB mass-loss have non-spherical morphologies.  Compact disks have been found in some bipolar PNe,  but their role in the shaping process remains unknown.  Compact Keplarian disks are found to be common around post-AGB binaries, however, 
these objects may never develop into PNe.  Another group of post-AGB stars, known as pre-PNe, are surrounded by collimated nebulae shining by reflected light or shock ionisation.  
We are observing the inner circumstellar regions of pre-PNe at high angular resolutions with the VLTI.  We seek to compare pre-PNe disks to those around other post-AGB stars and PNe.   New observations of the pre-PN, IRAS 16279-4757, show evidence for a disk similar to those seen in young PNe.    
\end{abstract}
\vspace{-0.1cm}
\firstsection 
\section{Introduction}
\vspace{-0.1cm}
At the end of the asymptotic giant branch (AGB) intermediate mass stars  go through an intense mass-loss phase when 50-90\% of their mass is expelled and expands in a nebula surrounding the core of the star.  After the loss of the envelope, the AGB star contracts and heats up, ionising the nebula ejected during the AGB phase, which shines as a planetary nebula (PN).  The reason for the dramatic increase in mass-loss during the late AGB phase is not well known.  AGB stars  appear to lose mass spherically, but the PNe that result from the mass-loss primarily have non-spherical morphologies. It is likely that the mechanism that causes the heavy AGB mass loss is related to the mechanism that dictates the ejecta's departure from sphericity.   Binarity may be an effective way to break spherical symmetry and to stimulate mass-loss.
 Some fundamental and rapid change in structure must occur between the spherical AGB mass-loss phase and the non-spherical, collimated PN phase (\cite[Iben 1995,  Balick and Frank 2002, De Marco 2009]{1995PhR...250....2I, 2002ARA&A..40..439B}).

Disks around evolved stars have been observed and are suspected to play a fundamental role in the shapes of PNe (e.g. Huggins 2007).
Theoretical work envisions their  role as the collimating agent for non-spherical mass-loss (e.g. Soker \& Livio 1994, Blackman et al. 2001).  
By studying the characteristics of these disks we can understand the engine that is forming the PNe.   Only with the Very Large Telescope Interferometer (VLTI) can we reach high enough spatial resolution to measure the parameters of the inner circumstellar environment of post-AGB stars,  in transition between the AGB and the PN phase. These parameters will meaningfully constrain single-star as well as binary models of AGB mass-loss and nebulae collimation. 
\vspace{-0.1cm}
\section{Previous Observations}
\vspace{-0.1cm}
\subsection{Compact disks around planetary nebulae central stars}
\vspace{-0.1cm}
Some PNe are found to harbour compact disks.  For example, disks have been observed with the VLTI in two young PNe, M2-9 and Mz 3 
(Lykou et al. 2011, Chesneau et al.  2007).   
These disks are compact toroids with an inner radius $\sim$ 10 AU, but contain much less mass  than their surrounding PNe, with the total dust mass in the disk $\sim$ 10$^{-5}$~M$_\odot$ 
(Lykou et al. 2011, Corradi et al 2011, Chesneau et al. 2007).  
The disks found in these PNe contain amorphous silicate grains, which implies a young age.  In addition, M2-9 has a binary companion with a period of $\sim$ 90 years (Lykou et al. 2011, Corradi et al. 2011).  Although a binary has not been detected in Mz 3, it is believed to exist with a similar period, due to its overall similarity to M2-9.  Finally, it must be noted that  both of these PNe may actually be PNe mimics 
(Frew and Parker et al. 2010).  

Another two young PNe, CPD-56$^\circ$8032 (\cite[De Marco et al. 2002]{2002ApJ...574L..83D}) and M2-29, also had disks detected in their cores. However, their properties are very different than those of M2-9 and Mz 3.  The compact disks have an inner radius $\sim$100 AU.  The disk mass is  $\sim$ 10$^{-3}$~M$_\odot$ for CPD-56$^\circ$8032 and  $\sim$ 10$^{-6}$~M$_\odot$ for M2-29 (Chesneau et al. 2006, Gesicki et al. 2010, Miszalski et al. 2011).  In addition, the disks found in these PNe both contain dual-dust chemistry (O and C rich). 
  No known binary exists for CPD-56$^\circ$8032, but a binary in M2-29 with a period of $\sim$ 17 years is likely (Hajduk et al. 2008).  

\vspace{-0.1cm}
\subsection{Compact disks around Post-AGB stars}
\vspace{-0.1cm}
Looking  to the post-AGB objects, which are closer on the evolutionary phase to when the disks were likely formed, we find two categories of objects:  1) naked post-AGB objects and 2) pre-PNe.    
\vspace{-0.1cm}
\subsubsection{Naked post-AGB objects}
\vspace{-0.1cm}
Naked post-AGB stars tend to not have a reflection nebulae, and as such are not thought to evolve into PNe (the exceptions are HD44179 and HR 4049, see Section 2.2.2).  While we call them ``naked" because they lack nebulosity, the SEDs of these stars actually show a larger IR-excess at near/mid infrared wavelengths, indicating the presence of a compact disk (De Ruyter et al. 2006).  Several of these disks have been observed with the VLTI and have an inner radius of $\sim$15 AU at 8 $\mu$m and with dust masses of $\sim$1x10$^{-2}$~M$_\odot$ 
(e.g. Deroo et al 2007, Deroo et al 2006).  
No expansion of the dust has been detected, implying that the dust is gravitationally bound in a Keplarian rotation.  The Keplerian rotation indicates that  it is  likely a relic of a strong interaction phase when the primary was an AGB giant and indicates that a binary companion to the post-AGB star is likely present. Such companions were eventually found with orbital periods between 100 and 2000 days 
(Van Winckel et al. 2009).  
Finally, the disks in the naked post-AGB stars exhibit strong crystalline silicate features.  This suggests that the disks found around naked post-AGB stars are actually older than the disks in some young PNe as the dust is more processed.
\vspace{-0.1cm}
\subsubsection{Post-AGB stars with a nebula (pre-PNe)}
\vspace{-0.1cm}
Some post-AGB stars, known as pre-PNe, have collimated nebulae shining in the optical by reflected light or shock ionisation  which are thought to become PNe.  It is unknown why some, otherwise similar, post-AGB stars have resolved nebulae while others do not.  Pre-PNe are very similar to young PNe and are extremely likely to be their immediate predecessors.  The SEDs observed for many pre-PNe are double peaked, indicative of a detached shell, others have a near-IR excess due to hot dust in the core (Lagadec et al. 2011).    Pre-PNe do not appear to harbour close binaries although wider binaries may be present (e.g.  IRAS 22272+5435 appears to have $>$ 22 year period, Hrivnak et al. 2011).  HD 44179  and HR 4049 are the only known pre-PNe that have compact disks detected.  However, they are unique in that they have both a naked post-AGB-style disk (Keplarian rotation, $\sim$10 AU in diameter,  with dual-dust chemistry, and a close binary companion with P $\sim$ 300 - 400 days; 
Men'shchikov et al. 2002, Dominik et al. 2003, Waelkens et al. 1991) and a reflection nebula, thereby classifying them as pre-PNe.  Therefore, we do not class HD 44179 or HR 4049 as  typical pre-PN, but as  ``cross-over" objects. 

We believe that the immediate circumstellar environments of pre-PNe, primarily the disks that are found there, hold the key to the breaking of AGB mass-loss symmetry which leads to collimated  PN morphology. The basic questions that drive our research are: What type of disks, if any, exist inside typical pre-PNe? What is the connection between the older, post-AGB Keplerian disks and the lower-mass, toroidal disks around young PNe? Are the disks a cause or a consequence of the asymmetric mass-loss?
  \vspace{-0.1cm}
\section{New and Future Observations with the VLTI}
\vspace{-0.1cm}
The VLTI has a spatial resolution up to 10 mas in the mid-IR (using MIDI) and up to 2 mas in the near-IR (using AMBER), making it a powerful tool to observe the inner circumstellar regions of post-AGB stars.  The VLTI allows us to determine the geometry and mass of a disk or any other structure present on the 10 -100~AU scale.  

We have carefully selected eight pre-PNe to observe with the VLTI.  Each target has a bright, compact, unresolved or slightly resolved (at the 0.3$''$ scale) core in the mid-IR as seen from VISIR observations (Lagadec et al. 2011).  Therefore, the chances of obtaining useful interferometric observations with the VLTI are very high.  All targets also have observed non-spherical nebulae around them.  We expect to find disks that have affected the shaping of the pre-PNe or that are a remnant of the shaping mechanism. Likely, these disks will have different characteristics from the naked post-AGB stars without a nebula.  Four of the eight pre-PNe have already been observed (from April - June 2011).  
Another four targets 
will be observed during November 2011 - April 2012.   

We report preliminary results on IRAS 16279-4757.  IRAS 16279-4757 is a complex axis-symmetric nebula located $\sim$ 2 kpc away, with intermediate inclination to the line of sight.  It is classified as a post-AGB object based on its SED (van der Veen, Habing and Geballe 1989).  Its optical spectrum suggests an inner star with spectral type of G5 (Hu et al. 1993).  It has PAHs and crystalline silicates similar to HD 44179 (Matsuura et al 2004).   We obtained  2 baselines of data with MIDI  
and 3 baselines with AMBER 

The resulting MIDI visibility curves show a sinusoidal pattern, indicating the presence of a ring-like disk (see Figure 1).  In addition, the visibility is high ($\sim$0.4 - 0.6) indicating the presence of a compact source in the centre.   Preliminary analysis of the visibility curves suggest the presence of a disk.  One baseline observed a disk with an inner radius of $\sim$ 150 AU, while the second baseline observed a disk with an inner radius of $\sim$ 70 AU.  This implies that the disk around IRAS 16279-4757 has an inner radius of $\sim$ 150 AU and we are viewing it at an inclination of $\sim$60$^\circ$ to the line of sight.  This size is similar to the disks seen in young PNe and larger than those observed around naked post-AGB stars.  Additionally, the sinusoidal pattern seen in the visibility curves is very similar to the visibility curves observed by the VLTI for the disk seen in CPD-56$^\circ$8032 (see Section 2.1) which has an inner radius of 97 $\pm$ 11 AU and an inclination of 28 $\pm$ 7$^\circ$ (Chesneau et al. 2006).   
The preliminary analysis of the AMBER observations indicates a 2-component source inside the disk found with the MIDI observations.  The 2 components are likely a stellar source and a possible second disk inside the first disk with a radius of $\sim$ 12 AU.  This second, smaller disk is similar in size to those around naked post-AGB stars. 

IRAS 16279-4757 has shown that promising results can be found by searching for disks inside pre-PNe with the VLTI. More detailed radiative transfer models will be conducted to derive disk  parameters such as inner and outer radii, scale-height, mass, and inclination.  These models will be similar to the models for Mz~3 and M2-9 (e.g. Lykou et al. 2010,  Chesneau et al. 2007).   In conclusion, observations with the VLTI coupled with detailed modelling of the disks will
lead to a quantified comparison between the physical properties of the disks which will allow us to study the evolutionary connection between the compact disks of naked post-AGBs and the tenuous, larger disks inside young PNe. 

\vspace{-0.1cm}
\begin{figure}
\begin{center}
 \includegraphics[width=4.2in]{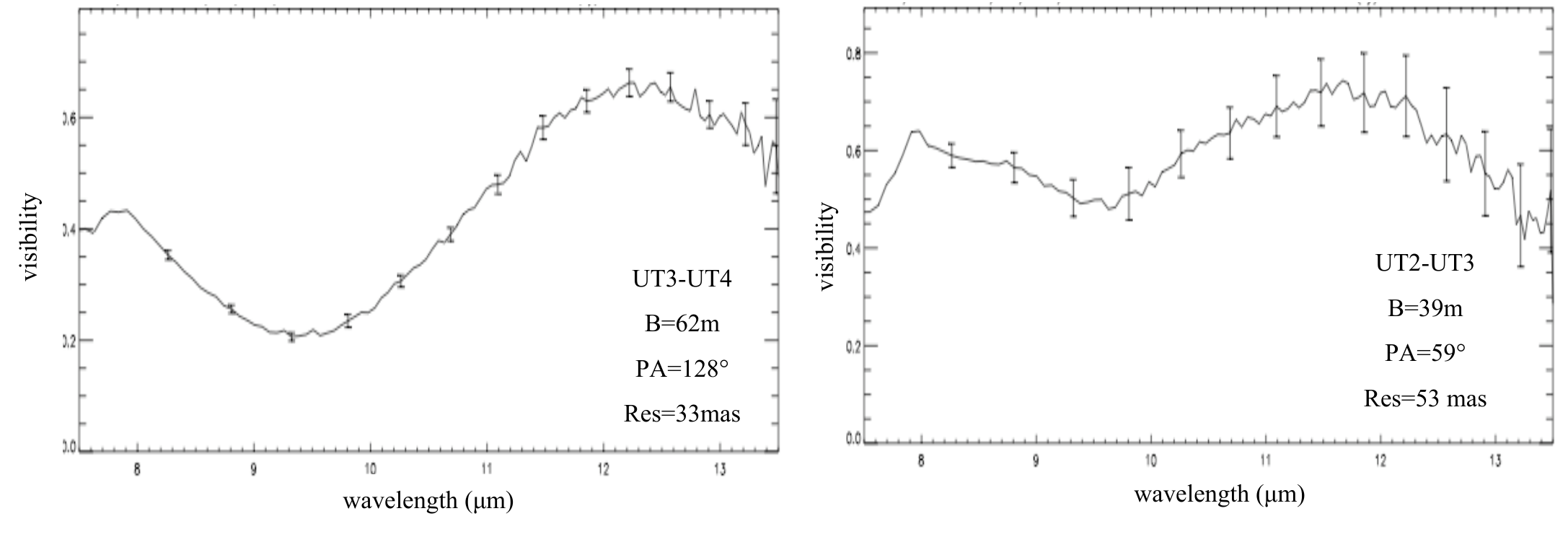} 
 \caption{MIDI visibility curves for IRAS 16279-4757 displaying a sinusoidal patter, indicating a ring-like disk.}   
   \label{fig1}
\end{center}
\end{figure}
\vspace{-0.1cm}

%


\end{document}